\documentclass[pra,twocolumn,showpacs,floatfix]{revtex4}
\usepackage{graphicx,amsmath,amsfonts,amssymb}
\usepackage[usenames]{color}

\begin{document}

\title{Matter-wave localization in a weakly perturbed optical lattice}

\author{Yongshan Cheng$^{1}$\footnote{yong\_shan@163.com}
and
   S. K. Adhikari$^2$\footnote{adhikari44@yahoo.com;
URL: www.ift.unesp.br/users/adhikari}}
\affiliation{$^1$Department of Physics, Hubei Normal University,
Huangshi 435002,
People's   Republic of  China\\
$^2$Instituto de F\'{\i}sica Te\'orica, UNESP - Universidade
Estadual Paulista,
%Barra Funda,
01.140-070 S\~ao Paulo, S\~ao Paulo, Brazil}

\begin{abstract} By numerical solution and variational approximation  of the
Gross-Pitaevskii equation, we studied the localization of a
noninteracting and weakly-interacting Bose-Einstein condensate in a
weakly perturbed optical lattice in one and three dimensions. The
perturbation  achieved through a  weak delocalizing expulsive or a
linear potential as well as a weak localizing harmonic potential
removes the periodicity  of the optical lattice and leads to
localization. We also studied some dynamics of the localized state
confirming its stability.

\end{abstract}


\maketitle

\section{Introduction}

Anderson suggested the possibility of localization of an electron
wave in a nonperiodic random potential about fifty years
ago \cite{anderson}. After the recent experimental realization of
Anderson localization of a Bose-Einstein condensate (BEC)
\cite{Nature-453-891,Nature-453-895}, this topic has been the
subject of intense research  activities. Billy {\it et al.}
\cite{Nature-453-891} observed the exponential tail of the spatial
density distribution of a $^{87}$Rb BEC after releasing it into a
one-dimensional (1D) waveguide with a controlled disorder potential
created by a laser speckle. Roati \emph{et al}.
\cite{Nature-453-895} observed the Anderson localization of a
noninteracting $^{39}$K BEC in a 1D quasiperiodic bichromatic
optical lattice (OL). {Recently, the experimental
realizations of three-dimensional (3D) localization of ultracold
gases of $^{40}$K \cite{SCI-334-66} and $^{87}$Rb \cite{11080137}
atoms in a 3D speckle potential were also reported.} In the
experimental studies of localization of a BEC,  bichromatic OL
\cite{PRL-98-130404,Nature-453-895}, shaken OL \cite{PRA-79-013611},
cold atom lattice \cite{cal} and speckle potentials
\cite{Nature-453-891} have been used. There have also been studies
of the effect of a repulsive atomic interaction on localization
\cite{experimental}. Bichromatic OL
\cite{adhikari,PRA-80-053606,quasiperiodic2}, cold atom lattice
\cite{PRL-95-020401,PRA-74-013616}, random potentials
\cite{NJP-10-045019,PRA-79-063604,random2,subdiffusive},
and OL plus harmonic trap \cite{quasiperiodic3},
among others, have been used in theoretical studies. Quasiperiodic
\cite{quasiperiodic1,quasiperiodic3} or random \cite{random1, PRL-95-070401,
PRL-95-170409,chapter3} potentials for Anderson localization were
realized by optical means. { In Ref. \cite{quasiperiodic3}, localization 
and Bloch oscillation of a BEC
was studied in two spatial dimensions for OL perturbed by a 
harmonic potential \cite{NATPHY-1-23}. Here we shall consider localization in 1D and 3D in the same potential.} 
%A bichromatic OL was obtained  by a
%primary OL perturbed by a secondary OL with an incommensurate
%wavelength \cite{Nature-453-895}.

However, there are other ways of obtaining a nonperiodic OL. Here we
{ consider another}
 approach for perturbing the OL, so that the
periodicity of the original OL is completely removed. If we perturb
the OL by a weakly delocalizing (i) linear potential or (ii)
expulsive harmonic potential, the periodicity of the OL is removed
without adding an extra confining term to the OL. Both linear
\cite{linear,SCI-282-1686} and expulsive \cite{expulsive} harmonic potentials can
be made in laboratory by optical and magnetic means.
% and hence are
% attractive from an experimental point of view. 
{ In Ref. \cite{SCI-282-1686}
 a tilted trap was used for the observation of Bloch
oscillations in a repulsive BEC, where  
 the external potential is
simply given by the sum of the linear (gravitational) component and
the OL laser field.  In Ref. \cite{expulsive} an expulsive potential was 
used to study solitons in BEC. }
%{ The gaussian beam profile
%of the lasers used to create the OL leads to 
% a residual harmonic trap, and hence can provide a
%weak harmonic confinement superimposed on the  OL \cite{NATPHY-1-23}.}

%For case (ii), in fact, the gaussian beam profile
%of the lasers $-$ a residual harmonic trapping potential $-$  leads to a
%weak harmonic confinement superimposed over the periodic potential.
%Thus the overall trapping configuration is inhomogeneous (see, e.g.,
%the recent review \cite{NATPHY-1-23}). In the realistic experimental
%setup, an OL is simply a set of standing wave lasers. The shape of
%the OL can be easily tailored by varying the parameters of the light
%field or applying an external magnetic field \cite{NATPHY-1-23}.}
%Compared with the secondary OL with an incommensurate wavelength in
%the bichromatic OL, a linear or expulsive perturbation is easily
%achieved.}

Here, with  the  numerical solution of the Gross-Pitaevskii (GP)
equation, we investigate the localization of cigar-shaped BECs in a
1D OL, weakly perturbed by linear and expulsive confining harmonic
potentials. We also considered adding a weak confining harmonic trap
to the OL, which removes the periodicity of the OL to obtain
localization. The localized states so obtained are completely
distinct and had a different spatial extension from the states
trapped in the confining harmonic trap alone and hence were
localized due to the nonperiodic OL. It is found that the localized
states in the three types of periodicity-broken OL are practically
the same and do not depend on the weak perturbation used to remove
the periodicity of the OL. In most cases the density profile of the
central part of the localized state is quite similar to a Gaussian
shape. Hence it is worthwhile to solve the GP equation analytically
using a Gaussian variational ansatz. Indeed we find that the
variational solution fits well the density distribution in the
central region, especially when the localization takes place
essentially on a single site of the perturbed OL. However,  the
localized states usually have a long undulating exponentially
decaying tail in either direction and extends  to several sites of
the OL corresponding to Anderson localization in a weak potential
\cite{chapter3,NJP-10-045019}. The stability of the localized state
under small perturbation is also investigated.

{We also extend our study to localization in 3D. In  3D, 
the effect of disorder on an initial state at large times 
depends strongly on the energy, and a mobility edge was predicted,
which corresponds to a critical energy at which a transition from
localized to extended character of the state 
occurs \cite{PNAS-69-1097}. As observed in the recent experiments in 3D
\cite{SCI-334-66, 11080137}, a two-component density distribution
emerges at large times 
consisting of an expanding mobile component and a
nondiffusing localized component 
after the ultracold gas  of atoms
are released into an optical speckle field. In this paper,
different from the experimental scenario, we focus our attention on the
stationary localized states, and not on the large-time expansion dynamics.} 
We find that the weakly perturbed 3D OL also leads to
localization. In the 3D realm, we only present results for the
expulsive perturbation of the OL, as other perturbations lead to
similar results. In this case we show results for noninteracting
BEC as well as for weakly repulsive BEC. The numerical chemical
potentials are in good agreement with the variational results in all
cases. The localized states of the 3D BEC are found to be stable
against small perturbations.

In Sec. \ref{II} we present the 1D and 3D models as well as their
variational analysis used in this study of localization.
In Sec. \ref{III} we study numerically the statics and dynamics of a
1D and a 3D BEC by analyzing the atom density profiles and chemical
potential of the localized states in a perturbed OL, and compare the
numerical results with the variational solutions. The study of a
sustained small oscillation of the localized states upon external
perturbation confirmed their stability. In Sec. \ref{IIII} we
present a summary of our findings.

\section{Analytical consideration }

\label{II}

The dynamic behavior of a weakly interacting Bose gas at low
temperature on a 3D OL is well described by the GP equation which has the form
of a (3+1)-dimensional nonlinear Schr\"odinger equation \cite{RMP-71-463}:
\begin{eqnarray}\label{eq3d}
i\frac{\partial \Phi}{\partial t}&=& - \frac{1}{2}\nabla^2\Phi
+G|\Phi|^2 \Phi+U(x,y,z)\Phi,\\
U(x,y,z)&=& -V_0[\cos(2x)+\cos(2y)+\cos(2z)], 
\end{eqnarray}
where $\Phi\equiv\Phi(x,y,z,t)$ is the BEC wave function with
normalization $\int \int \int dx dy dz |\Phi|^2 =1$, and $\nabla^2$
is the 3D Laplacian. The constant $G\equiv4\pi aN$ characterizes the
two-body interaction, $N$ the total number of atoms,    $a$ the
atomic scattering length, and $U$ 
the 3D OL.  The spatial variable, time, density
$|\Phi(x,y,z, t)|^2$, and potential are measured in units of
$l_0=\lambda/(2\pi)$, $ml_0^2/\hbar$, $l_0^{-3}$ and
%$\hbar^2/(ml_0^2)$, 
$2E_R$,
respectively, where $E_R=h^2/(2m\lambda^2)$ is the recoil energy of an atom absorbing a 
lattice photon,  $\lambda$ is the wavelength
of the OL potential, and $m$ is the atomic mass.

We consider a 3D trapping potential $U(x,y,z)$ made up of three
weakly perturbed OL components defined as:
\begin{eqnarray}\label{pot3d}
U(x,y,z) =\sum_{i=1}^3V(x_i),
\end{eqnarray}
where $x_{1,2,3}$ correspond to $x,y,z$, respectively. Each of the
three components has the following   forms:
\begin{eqnarray}\label{pot1}
V(x_i) = -V_0\cos(2x_i)-C x_i,  \\
V(x_i) = -V_0 \cos(2x_i)\pm C x_i^2,   \label{pot2}
\end{eqnarray}
with $V_0=5$,
corresponding to perturbations by a weak linear [Eq. (\ref{pot1})]
and confining [Eq. (\ref{pot2}), positive sign] and expulsive [Eq.
(\ref{pot2}), negative sign] traps, with strength $C$. These
perturbations slightly modify the periodic OL and generate
nonperiodic OL potentials.
%In the realistic experimental setup, a
%periodic OL potential is created by two or more counter-propagating
%laser beams. The lattice depth $V_0$ and wavenumber $k$ can be
%varied at will.
% In this present work, in rescaled units, we let $V_0
%= 5,\ k=2$ which are experimentally accessible. With a weakly
%expulsive perturbation, the 3D OL potential Eqs. (\ref{pot3d}) is
%shown in Fig. \ref{fig1}(a) (slice plot) and 1(b) (contour plot with
%$U=-7$).

\begin{figure}%[!ht]
\begin{center}
\includegraphics[width=\linewidth]{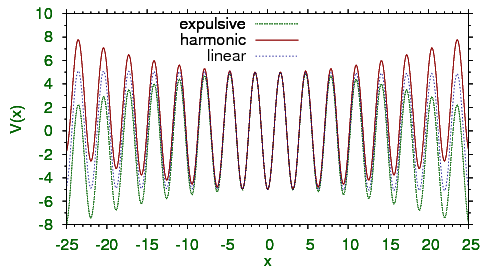}
\end{center}

\caption{(Color online) The linear, confining, and expulsive
harmonic potentials $V (x)$ as a function of $x$ in dimensionless
units from Eqs. (\ref{pot1}) and (\ref{pot2}) for $C = 0.005, V_0=5$.
 } \label{fig1}
\end{figure}

We also consider a cigar-shaped condensate oriented in the
longitudinal direction ($x$ direction). The trapping potential is a
weakly perturbed OL in the longitudinal direction and the transverse
($y, z$ directions) dynamics of the condensate is frozen to the
respective ground states of harmonic traps. This cigar-shaped system will
be described by using an effective 1D model derived from the 3D GP
equation (\ref{eq3d}) by integrating over the transverse variables
\cite{1d}. Then, the time-dependent wave function
$\phi\equiv \phi(x,t)$ of a BEC can be described by the following
 1D GP equation \cite{1d,1dd}
\begin{eqnarray}\label{eq1}
i\frac{\partial \phi}{\partial t}=- \frac{1}{2}\frac{\partial^2
\phi}{\partial x^2} +g|\phi|^2 \phi+V(x)\phi,
\end{eqnarray}
with normalization $\int_{-\infty}^{\infty}|\phi|^2 dx=1$,
nonlinearity $g= 2aN/a_\perp^{2}$, and potential $V(x)$
given by Eq. (\ref{pot1}) or (\ref{pot2}). The spatial variable $x$,
time $t$,
density $|\phi|^2$,
and energy are expressed in normalized  units
$a_\perp=\sqrt{\hbar/(m\omega_\perp)}$, $\omega_\perp^{-1}$, $a_\perp^{-1}$
 and
$\hbar\omega_\perp$, where $\omega_\perp$ is the angular frequency
of the transverse trap.
 These weakly perturbed OLs along $x$ axis are
illustrated in Fig. \ref{fig1} which shows that the OL dominates
for small $x$, but the periodicity of the original OL is completely
removed.

 Stationary solutions $u(x,y,z)$
to Eq. (\ref{eq3d}) can be given by $\Phi(x,y,z, t)=\exp(-i\mu
t)u(x,y,z)$, with $\mu$ the chemical potential. For convenience, we
use unified notation $u$ to denote the real wave function with
$u\equiv u(x)$ for 1D and $u\equiv u(x, y ,z)$ for 3D BEC,
respectively. We choose $d (= 1,3)$ to denote the dimension of
space. The stationary real wave function obeys
\begin{eqnarray}\label{eq13d}
\mu u=- \frac{1}{2}\sum_{i=1}^d \frac{\partial^2 u}{\partial x_i^2}
+G u^3+ \sum_{i=1}^d V(x_i)u.
\end{eqnarray}
The nonlinearity $G$ should be replaced by $g$ in 1D.
Equation (\ref{eq13d}) can be derived from the Lagrangian
density
\begin{eqnarray}\label{den}
{\cal L}=\mu u^2-\frac{1}{2}\sum_{i=1}^d \left(\frac{\partial
u}{\partial x_i}\right)^2 -G \frac{u ^4}{2} -\sum_{i=1}^d V(x_i)
u^2.
\end{eqnarray}
To apply the variational
approximation to Eq. (\ref{eq13d}), we use the variational
Gaussian ansatz:
\begin{eqnarray}\label{ANS3d}
u=\left(\frac{1}{\pi^{1/4}\sqrt w}\right)^d\sqrt{ {{\cal N}} }
\prod_{i=1}^d \exp\left(-\frac{ x_i^2} {2w^2}\right) ,
\end{eqnarray}
with $w$ the width and ${\cal N}$  the normalization of the
localized BEC. The effective Lagrangian of the system can be found
upon by substituting Eq. (\ref{ANS3d}) into Eq. (\ref{den}) and
integrating over space \cite{CYS}:
\begin{eqnarray}\label{LAG3d}
L&=& \mu {\cal N}- \mu-\frac{d{\cal N}}{4w^2}- \frac{G{\cal
N}^2}{2}   \left(\frac{1}{\sqrt{2\pi}w}\right)^d+ dV_0 {\cal N} e^{-w^2}\nonumber \\
& -& Rd \frac{  {\cal N}w^2}{2},
\end{eqnarray}
where $R=0$ for the linear potential and $R=\pm C$ for the confining
and expulsive harmonic potential. The Euler-Lagrange equation
$\partial L/\partial \mu =0$ yields the normalization ${\cal N}=1$.
We  use it in the following equations. The remaining equations
$\partial L/\partial w = \partial L/\partial {\cal N}=0$ yield,
respectively,
\begin{eqnarray}\label{WID3d}
1&=&-\frac{G}{(2\pi)^{d/2} w^{d-2}}+ 4 V_0 w^4 \exp(-w^2)+ 2Rw^4  ,\\
\mu&=&\frac{d}{4w^2}+ \frac{G}{(w\sqrt{2\pi})^d}  -dV_0
\exp(-w^2)+dR\frac{w^2}{2}, \label{CHE3d}
\end{eqnarray}
and determine the width $w$ and the chemical potential $\mu$.
%This
%set of equations leads to  a straightforward conclusion that
The linear
perturbation has no effect on the width and chemical potential of
the stationary localized states because $R=0$ for the
 linear perturbation.

\section{Numerical Results}

\label{III}

We performed the numerical integration of GP equations (\ref{eq3d})
and (\ref{eq1}) employing the imaginary or real-time split-step
Fourier spectral method with space step $0.04$, time step $0.001$.
We checked the accuracy of the results by varying the space and time
steps and the total number of space and time steps. We also checked
the results with those obtained from a split-step Crank-Nicolson
algorithm using same steps \cite{CPC}.

By numerically solving equation (\ref{eq1}), we first investigate
the effects of the weakly perturbed OL on a localized state in 1D
BEC. To obtain a stationary localized state, the initial input pulse
is taken as $\phi(x,0) =\exp(-x^2/2) /\pi^{1/4}$ with a parabolic
trap $V'(x)=x^2/2$ in our numerical integration. During the
numerical simulation, the parabolic trap is slowly turned off, and
the potential (\ref {pot1}) or (\ref{pot2}) and the nonlinearity $g$
is slowly turned on with a very small increment ($=0.00001$) in each
time step. The time evolution is continued until the parabolic trap
is completely turned off and the potential (\ref {pot1}) or (\ref
{pot2}) is completely turned on, and the the nonlinearity $g$ has
the desired value.

\begin{figure}%[!ht]
\begin{center}
\includegraphics[width=.49\linewidth]{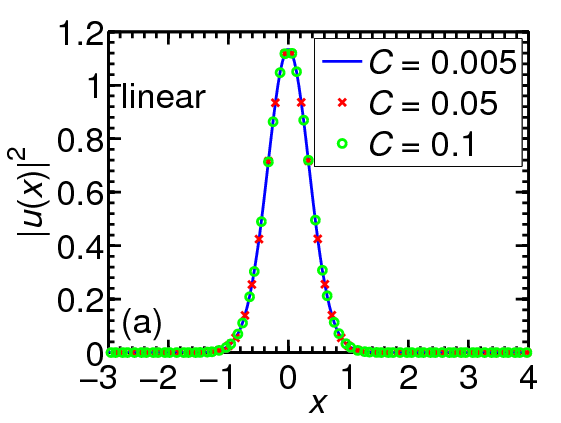}
\includegraphics[width=.49\linewidth]{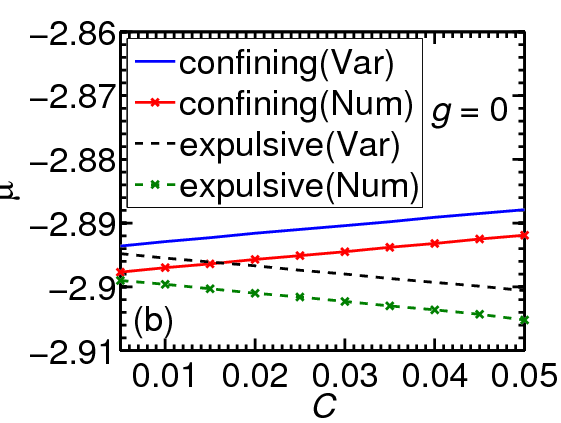}
\end{center}

\caption{(Color online) (a) Normalized numerical density profile
$|u(x)|^2$ of the localized BEC versus normalized $x$ for potential
(\ref{pot1}) with $C=0.1,0.05, $ and 0.005 and $g=0$. (b) Normalized
numerical (Num) and variational (Var) chemical potentials $\mu$ of
the localized BEC with $g=0$ versus $C$ for the confining and
expulsive perturbed OL (\ref{pot2}). } \label{fig2}
\end{figure}

In Figs. \ref{fig2} we exhibit typical numerical and variational
results of the localized states with different perturbations for
$g=0$. In the case of the linearly perturbed OL  [Eq. (\ref{pot1})],
Fig. \ref{fig2} (a) shows that the numerical density profiles
{ of the central part} of the localized states are the
same for $C=0.005, 0.05$ and 0.1, and the numerical chemical
potentials are all $-2.897$ in agreement with the variational
result, $-2.894$. For confining and expulsive perturbed OL
(\ref{pot2}), in Fig. \ref{fig2} (b) we plot the numerical and
variational chemical potentials $\mu$ of the localized states. In
the case of the expulsive perturbation, the chemical potential
decreases with the perturbation  strength. The opposite happens for
the confining perturbation. The reason is that an expulsive
perturbation weakens the trapping of the OL and a confining
perturbation strengthens the trapping.  For small $C$, the influence
of $C$ on the density profile and chemical potential of a localized
state is small. We also find  that the stability of the localized
states is related to the perturbation strength $C$. If we increase
$C$, the stability situation does not improve. But if we reduce $C$
the localized states could be unstable as $C\to 0$. So we keep
$C\geq0.005$ in this paper.

\begin{figure}%[!ht]
\begin{center}
\includegraphics[width=.49\linewidth]{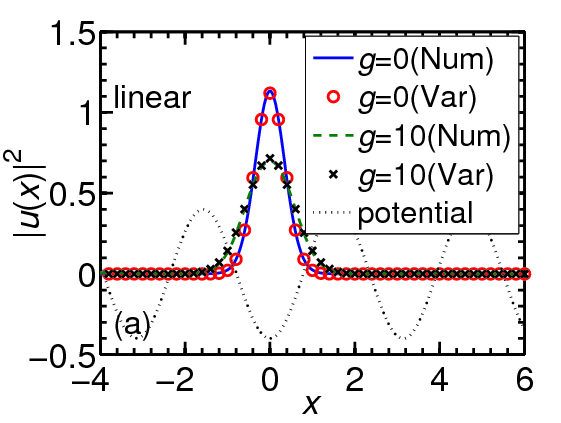}
\includegraphics[width=.49\linewidth]{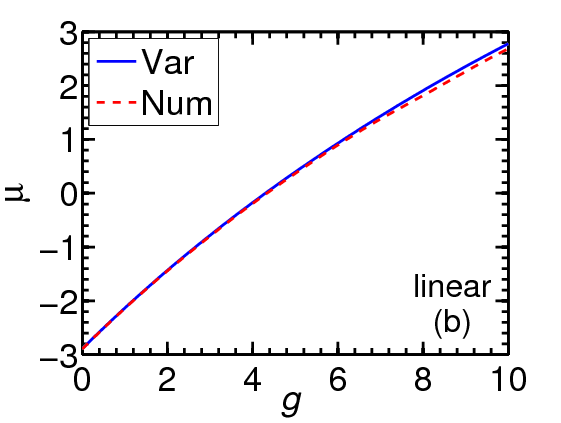}
\includegraphics[width=.49\linewidth]{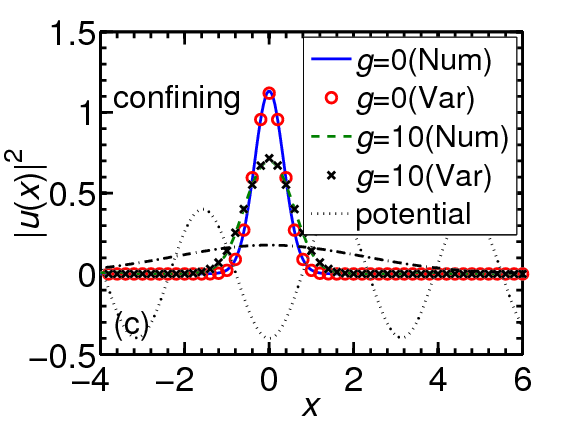}
\includegraphics[width=.49\linewidth]{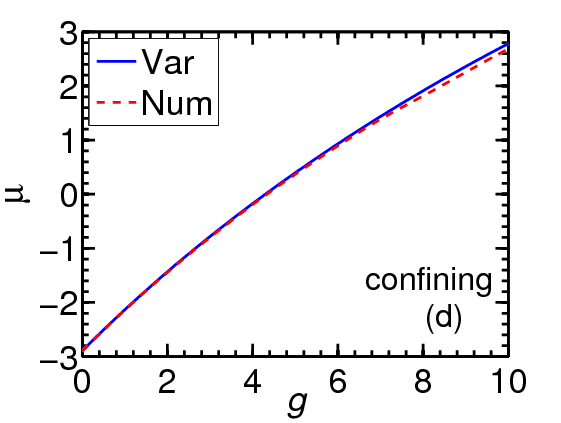}
\end{center}

\caption{(Color online)   Numerical (Num) and variational (Var)
 density profile $|u(x)|^2$ of the localized BEC  versus $x$ for
the (a) linear and   (c) confining harmonic perturbations of the OL
with $C=0.005$,  and $g = 0$ and 10, respectively. In (c), we also
show the density profile of the localized BEC trapped by only the
confining harmonic potential $V(x)=0.005 x^2$ with $g=0$ (the black
dot-dashed line). The numerical (Num) and variational (Var) chemical
potentials $\mu$ of the localized BEC  versus $g$ for (b) linear and
(d) confining perturbations of OL. All quantities are dimensionless.}
 \label{fig3}
\end{figure}

With the periodic potential  $-5\cos(2x)$, the noninteracting
Schr\"{o}dinger equation [$viz. \ g=0$ in Eq. (\ref{eq1})] permits
only delocalized states in the form of Bloch waves. Localization is
possible in the noninteracting Schr\"{o}dinger equation due to the
nonperiodic nature of the perturbed OL, as shown in Figs. \ref{fig3}
(a) and  (c) where we exhibit the typical density profiles of the
localized states in the linear and confining harmonic  potentials
for $g=0$ and 10 and compare them with the corresponding variational
results.  To confirm that these localized states are induced by the
weak perturbation of the  OL, in Fig. \ref{fig3} (c), we also show
the localized state for $g=0$ for the confining harmonic potential
alone (see the black dot-dashed line). This state has a different
spatial extension and can be distinguished from the other states
localized due to the perturbation of the OL. Figures \ref{fig3} (b)
and (d) show, respectively, the effect of the nonlinearity on the
chemical potential (again in the agreement with the prediction of
the variational results). The corresponding results for the
expulsive perturbed potential are in agreement with  those of the
confining potential, and are not shown here. { To
understand the effects of $g$ on $\mu$, and on localization in general, 
it is useful to consider the variational equations (\ref{WID3d}) and 
  (\ref{CHE3d}); for $g=0$, $\mu =-2.894$. 
As $g$ is increased, $\mu$ increases and the system is more repulsive
and extends over more region in space 
and the localization length increases. Eventually, at $g\approx 18.6$ 
the state is completely delocalized and beyond this value of $g$, 
no localization is possible and there is no real solution of Eq. (\ref{WID3d})
for the width.}

%it is helpful to revisit the
%Eq. (\ref{CHE3d}). The first term on the right-hand side of Eq.
%(\ref{CHE3d}) corresponds to the quantum kinetic energy coming from
%the uncertainty principle; the second term is induced by the
%mean-field interaction(the double mean-field interaction energy);
%and the last two terms with a total negative energy arise from the
%perturbed OL that contributes an effective potential well to the BEC
%system. The effective potential well results in the negative
%excitation energy. As the nonlinearity $g$ increases, the positive
%mean-field interaction energy increases, and the chemical potential
%increases. Because the effect of $g$ on the width $w$ [see the Eq.
%(\ref{WID3d})], the chemical potential is nonlinearly associated
%with $g$. When $g$ increases to a critical value, the chemical
%potential balances. As $g$ increases from the critical value, the
%sign of the total chemical potential changes from a negative to
%positive value which is indicated in Figs.\ref{fig3} (b) and (d).
%When $g$ is large enough, of course, the BEC system becomes more
%repulsive and hence spreads beyond a single OL site. When $g>18.6$,
%it is also shown by Eq. (\ref{WID3d}) that the variational width has
%not a finite real solution for $V_0=5$ and $C=0.005$. For accuracy,
%we limit $g\leq10$ in the all numerical calculations. }

\begin{figure}%[!ht]
\begin{center}
\includegraphics[width=\linewidth]{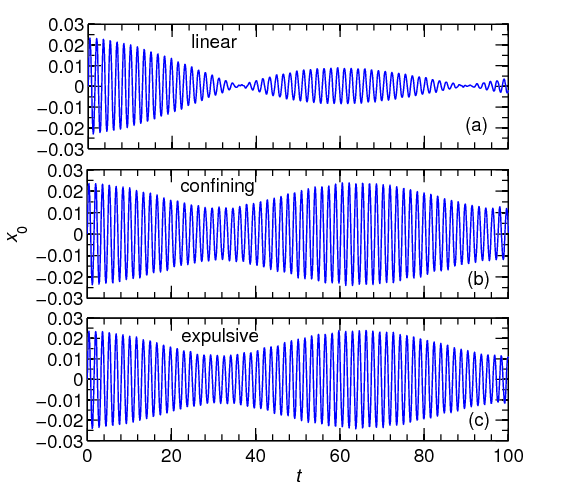}
\end{center}

\caption{(Color online) Center of the localized state
$x_0$ versus  time $t$ during the oscillation of the
localized BEC on the nonperiodic OL (\ref{pot1}) and (\ref{pot2})
initiated suddenly by introducing an initial momentum $p_0=0.1$
through the transformation $u(x)\to u_0(x)\exp(ip_0x)$ with $g=0$
and $C=0.005$ for the (a) linear,  (b) confining, and (c) expulsive
perturbation, respectively. } \label{fig4}
\end{figure}

As a typical Hamiltonian system, the stability of an ordinary
localized BEC can be determined by the Vakhitov-Kolokolov (VK)
criterion \cite{VK-criterion}. Of course, the direct applicability
of this criterion depends on the form of the nonlinearity and on the
form of trapping potential \cite{PhysRep-303-259}. In the case of an
attractive BEC confined in a magnetic trap, for example, a general
stability criterion is derived for the ground states of the GP
equation in Ref. \cite{PRA-62-023607}. This criterion states that a
localized state is stable only if the chemical potential $\mu$ and
the number of atoms $N=\int_{-\infty}^{\infty}|u(x)|^2 dx$ satisfy
$\partial N/\partial\mu < 0$\cite{PRA-62-023607}. The
chemical potential $\mu$
corresponds to $-\Lambda$ of Ref. \cite{PRA-62-023607}. As for the
repulsive nonlinearity, however, the stability of gap soliton
families obeys an inverted (``anti-VK") criterion $\partial
N/\partial\mu > 0$ \cite{PRA-81-013624}. {In Eq.
(\ref{eq1}), the number of atoms $N$ is linearly
related  to the nonlinearity via $g= 2aN/a_\perp^2$, and the wave
function is normalized $\int_{-\infty}^{\infty}|u(x)|^2 dx=1$.}
Then, the anti-VK criterion can be expressed as $\partial
g/\partial\mu > 0$. Figures \ref{fig3} (b) and 3(d) show that the
anti-VK criterion $\partial g/\partial\mu > 0$ holds, hence the
localized states are stable in the parameter range. To justify the
anti-VK criterion, the stability of the localized BECs is tested in
systematic numerical integration of GP equation (\ref{eq1}) with
$g>0$. For example, with $g = 10$, first we create a   localized BEC
in a nonperiodic OL. Then, the stability of the localized states was
established by slightly changing the potential (by multiplying by a
factor of 0.98) in the real-time routine and continuing the time
evolution to 10000 units of time. We find  that the localized states
remain localized in real-time routine and hence are considered to be
stable. {  At large time scales, the behavior
of an initially localized wave packet has been analyzed numerically
in Ref. \cite{subdiffusive}.  It was shown that above a certain
critical strength of nonlinearity a subdiffusion occurs.  We
emphasize here that, in our numerical calculations, the stationary
localized states are obtained in an adiabatic way, and the stability
of the localized states was tested within a short period of time.}

Nevertheless, the anti-VK criterion can not be naturally extended to
the noninteracting BEC ($viz. \ g=0$). To test the stability of
these noninteracting localized BEC, first we create a localized BEC
in a nonperiodic OL. Successively, at $t=0$, we suddenly introduce a
small initial momentum $p_0=0.1$ by changing  $u(x)\to
u_0(x)\exp(ip_0x)$, where $u_0(x)$ is the wave function of the
localized BEC at $t=0$. The center $x_0$ of the localized states
versus time $t$ is presented in Figs.  \ref{fig4}  where $x_0$ is
obtained by instant Gaussian-function fitting to the central region
of the density distribution. It is found that after the
perturbation, the localized states perform a sustained oscillations
but remain localized which imply that the localized states are
stable against  small perturbations. Contrary to our intuition,
however, the oscillations are quasiperiodic because of the
deformation of the density profile, as explained in Ref.
\cite{PRA-83-023620}.

\begin{figure}%[!ht]
\begin{center}
\includegraphics[width=\linewidth]{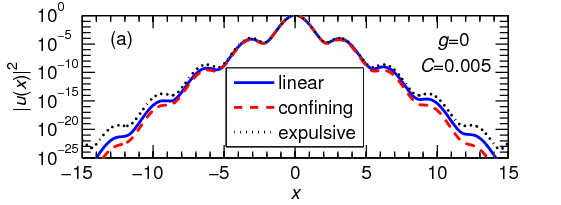}
\includegraphics[width=\linewidth]{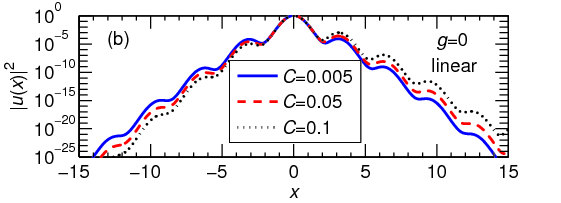}
\end{center}

\caption{(Color online)  (a) Normalized  numerical  density profile
$|u(x)|^2$ of the localized BEC versus $x$ with $g=0$ and $C=0.005$
for the linear, confining and expulsive perturbations, respectively.
(b) Normalized  numerical  density profile $|u(x)|^2$ of the
localized BEC versus $x$ with $g=0$ and $C=0.005, 0.05, 0.1$ for the
linear perturbations.} \label{fig5}
\end{figure}

One interesting earmark of Anderson localization in a weak
disordered potential is a long exponential tail of the localized
state inspite of the Gaussian distribution of the central part
\cite{RepProgPhys-73-062401}. To observe the effect of different
perturbations of the OL on the tail of localized state, we present
in Fig. \ref{fig5} the density profiles of the localized BEC on log
scale. As shown in Fig. \ref{fig5} (a), the localized BECs in the
three perturbed OLs have long exponential tails extending far beyond
the central Gaussian distribution. { Further
investigations show that the exponential tails are symmetric with respect to 
$x=0$ for the confining and expulsive perturbations,
i.e., both left and right localization lengths are the same.
In these cases, the localization length decreases with the increase of $C$.
For linear perturbation, both the perturbed OL and the exponential tails are
asymmetric with respect to $x=0$, and the asymmetry increases with the increase of the
perturbation strength $C$, as shown in Fig. \ref{fig5} (b). Although,
the localization lengths  depend on the
perturbation strength $C$, the dependence is weak for small $C$.}

\begin{figure}%[!ht]
\begin{center}
\includegraphics[width=.49\linewidth, clip]{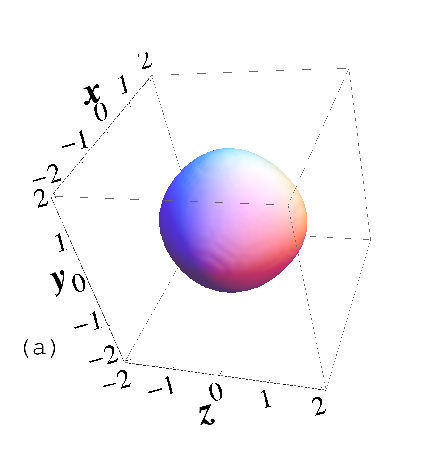}
\includegraphics[width=.49\linewidth, clip]{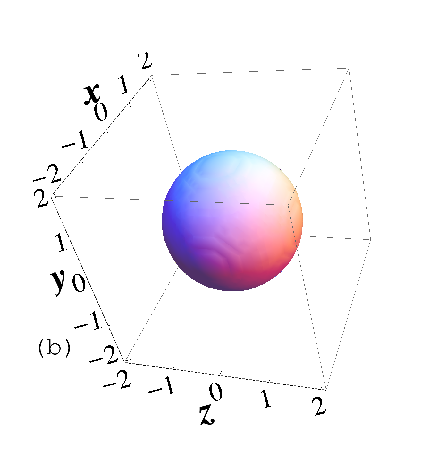}
\end{center}

\caption{(Color online)   The 3D contour plot of normalized density
profile of the localized state for $G=0$ from (a) numerical
calculation and (b) variational approximation with the 3D OL
perturbed by the expulsive potential (\ref{pot2}) with $C=0.005$.
In the plot
the contour was set at a density of 0.001.
  } \label{fig6}
\end{figure}

\begin{figure}%[!ht]
\begin{center}
\includegraphics[width=.49\linewidth, clip]{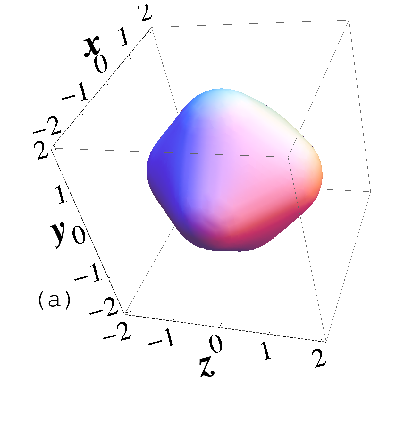}
\includegraphics[width=.49\linewidth, clip]{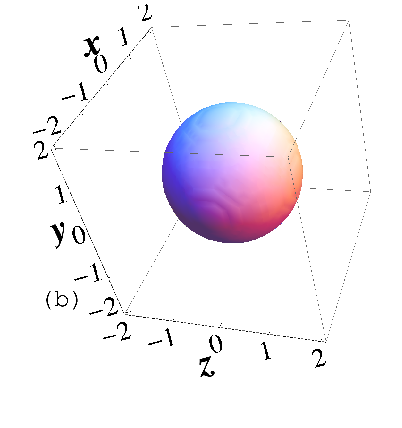}
\includegraphics[width=.49\linewidth, clip]{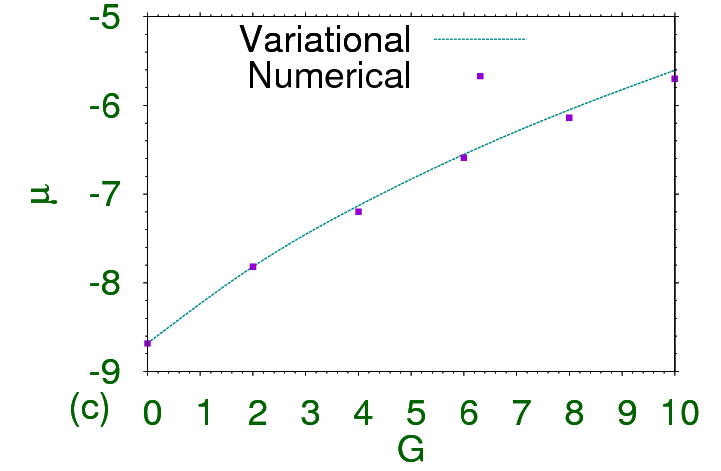}
\includegraphics[width=.49\linewidth, clip]{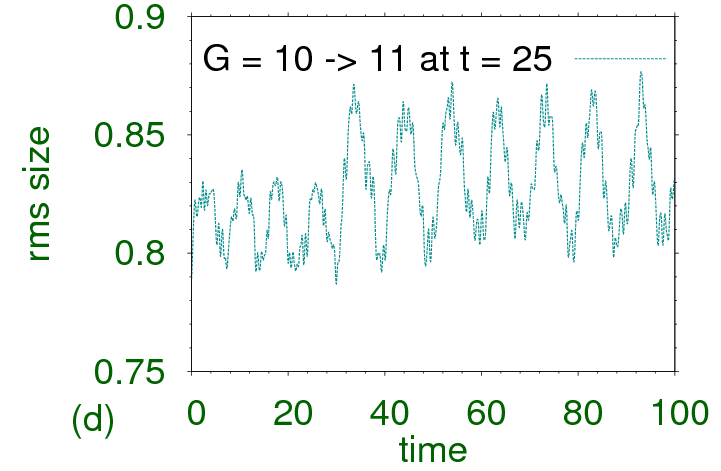}
\end{center}

\caption{(Color online)  (a) and (b) The same as Figs. 6 for $G=10$.
(c) Chemical potential $\mu$ for different $G$ from
numerical simulation and variational approximation. (d) Width
$w$ for $G=10$ during real time evolution when at $t=25$ the
nonlinearity $G$ was suddenly changed from 10 to 11.  } \label{fig7}
\end{figure}

Now we study localization in the 3D system with a perturbed OL Eq.
(\ref{pot3d}). As in a 1D system, the results  for three types of
perturbations are very similar in the 3D system. Hence, in a 3D
system we illustrate the results for the expulsive perturbation
only. First, we illustrate the results for the noninteracting case:
$G=0$. To test the stability of the localized state for $G=0$ we use
the real-time propagation of the 3D GP equation (\ref{eq3d}) with
the initial state at $t=0$ taken as the variational localized state.
The initial solution settles down to the final localized state after
iteration during an adequate interval of time. The 3D contour plot
of the density profiles of the localized state with its initial
variational counterpart are shown in Figs. \ref{fig6} (a) and (b).
Although the variational solution is spherically symmetric, this
symmetry is absent in the actual localized state.  This is because
the 3D perturbed OL is not spherically symmetric.
In spite of this, the variational
approximation is very useful in the study of localization, and the
numerical solution is nearly spherically symmetric. The variational
chemical potential is $-8.6826$ compared to the numerical result of
$-8.690$.

Next, we study 3D localization in a repulsive system with $G=10$.
Numerically, we find the stable localized state by real-time
propagation with the variational approximation as the initial input.
The 3D contour plots of the localized state with its  variational
counterpart are shown in Figs. \ref{fig7} (a) and (b). For $G=10,$
the numerical profile of the localized state in Fig. \ref{fig7} (a)
shows more deviation from spherical symmetry compared to \ref{fig6}
(a) for $G=0$. The variational chemical potential of the localized
state in Fig. \ref{fig7} (a) is $-5.604$ compared to the numerical
result of $-5.65$. In Fig. \ref{fig7} (c) we show the numerical
chemical potential of the localized states for different $G$ and
compare that with the variational results. Finally, we test the
stability of the localized state with $G=10$. While obtaining the
localized state, during real-time propagation we suddenly change $G$
from 10 to 11, making the system more repulsive. Consequently, the
root mean square size $w$ increases and the system executes
breathing oscillation. In Fig. \ref{fig7} (d) we plot the root mean
square size of the system during time evolution when $G$ was
increased at $t=25$. The sustained oscillation of the system
demonstrates the stability of the localized state.
{ The variational equation (\ref{WID3d}) gives the largest value of $G$ 
for which there is a localized state. Real solutions for the width 
are allowed for $G<230.7$, beyond which there is no localization. }

\section{Summary and Discussion}
\label{IIII}

We suggested several simple schemes for removing the periodicity of
an OL by adding a weak nonperiodic perturbation to study Anderson
localization. We suggested three perturbations in the form of
linear, confining harmonic and expulsive potentials. All these
perturbing potentials can be  created in laboratory and such
perturbed lattices can easily be used for BEC study. Usually a
bichromatic lattice is used in BEC studies, where the periodicity of
the principal OL is removed by adding a secondary OL of
incommensurate wave length.  
%It is more difficult to make such a
%bichromatic lattice in laboratory than the present perturbed
%nonperiodic OL.

Using the weakly perturbed OL we studied the problem of localization
of a noninteracting or weakly interacting BEC in 1D and 3D systems
using numerical and variational approaches to the GP equation. We
studied the density profile and chemical potential of the localized
states, where good agreement between variational and numerical
results was established.  The stability of the localized state was
established by the Vakhitov-Kolokolov  criterion and complemented by
numerical studies of stable oscillation of the localized state upon
small perturbation.

{ In this study we were mostly concerned with {\it stationary} 
localized states 
of BEC in 1D and 3D in nonperiodic potential and their stability. We are 
not concerned with the very long-time dynamics of an {\it arbitrary} initial state 
in such a potential. For example,  in a 1D weakly nonlinear disordered lattice
Anderson localization 
is destroyed and the field spreads subdiffusively far 
beyond the linear localization zone at large times \cite{subdiffusive}. Also, in 3D if a 
wave packet propagates in a linear disordered potential, at intermediate times 
there is a diffusion regime (due to atoms that have energies above the mobility 
edge), and at long times, if the disorder is strong enough, a remaining localized 
part (due to atoms that have energies below the mobility edge) will emerge
\cite{PNAS-69-1097,SCI-334-66}.   
Nevertheless, we demonstrate here using a mean-field model 
that stationary localized states are possible 
in a disordered potential in 1D and 3D for a relatively large nonlinearity 
or atomic interaction and this could motivate new experiments in the localization of
BEC. 
}

\acknowledgments
%\ack

FAPESP and CNPq (Brazil) provided partial support. Y.  Cheng
undertook this work with the Science and Technology Program of the
Education Department of Hubei, China, under Grants No. D200722003
and Z200722001.

\hskip .2 cm

%\section*{References}

\end{document}